# Bond Breaking Kinetics in Mechanically Controlled Break Junction Experiments: A Bayesian Approach

*Dylan Dyer[1] and Oliver L.A. Monti[1,2]\**

[1] Department of Chemistry and Biochemistry, The University of Arizona, Tucson, Arizona 85721, United States

[2] Department of Physics, The University of Arizona, Tucson, Arizona 85721, United States

\* Phone: +520 626 1177, Email: monti@arizona.edu

Breakjunction experiments allow investigating electronic and spintronic properties at the atomic and molecular scale. These experiments generate by their very nature broad and asymmetric distributions of the observables of interest, and thus a full statistical interpretation is warranted. We show here that understanding the complete distribution is essential for obtaining reliable estimates. We demonstrate this for Au atomic point contacts, where by adopting Bayesian reasoning we can reliably estimate the distance to the transition state, $x^\ddagger$, the associated free energy barrier, $\Delta G^\ddagger$, and the curvature $v$ of the free energy surface. Obtaining robust estimates requires less experimental effort than with previous methods, fewer assumptions, and thus leads to a significant reassessment of the kinetic parameters in this paradigmatic atomic-scale structure. Our proposed Bayesian reasoning offers a powerful and general approach when interpreting inherently stochastic data that yield broad, asymmetric distributions for which analytical models of the distribution may be developed.

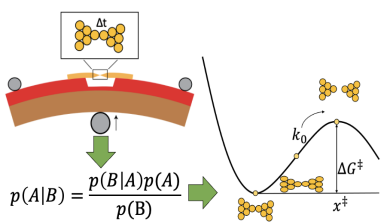

**molecular electronics, breakjunction experiment, Au nanowires, speed dependence, distributed data**

In single molecule electronics, test beds have been developed to isolate single organic molecules and to characterize their conductance, $G = 1/R$. Charge transport at the nanoscale deviates significantly from Ohm's law, since macroscopic diffusive transport shifts into microscopic ballistic transport when length scales are comparable to the de Broglie wavelength. This requires the ballistic transport to be treated in the fully quantum regime, which is probabilistic by nature. Consequently, measurements performed in either Mechanically Controlled Break Junction (MCBJ) or Scanning Tunneling Microscope-Break Junction (STM-BJ) experiments yield widely distributed and asymmetric conductance and lifetime histograms for both pure Au and molecular junctions. How to interpret these distributions and reliably extract the most information is an ongoing challenge, since these distributions may contain valuable yet thus far untapped information on the physics of quantum transport in such atomic-scale contacts. Worse, the distributions may obscure the true underlying physics, leading to biased or potentially even wrong conclusions. This prompts the need for microscopic insights into the factors that contribute e.g. to the broad experimental conductance or lifetime histograms, with the aim to invert the distributions into atomistic behavior and properties. This is the fundamental challenge in break junction experiments, where data is usually widely distributed.

Progress has been made to develop analytical expressions for the conductance histogram in terms of parameters that describe the kinetic profile of the junction and its mechanical manipulation, including the histogram shape. This shape contains information on the level alignment of the channel and the channel-electrode couplings.[1,2] Despite these advances in generating possible distributions based on microscopic parameters, inverting experimental distributions to retrieve the underlying model parameters remains a formidable task. The situation is worse for extracting information from lifetime histograms because this has mostly been limited to single point measures such as the mean or most likely value.

Here we show that kinetic parameters for bond breaking obtained in this manner needs to be reassessed. To achieve this, we employ an effective one-dimensional fully microscopic model developed by Dudko et al. for force experiments on biomolecules to capture all relevant kinetic aspects of bond breaking in the junction.[3] Inverting the experimental lifetime distributions allows us thus for the first time with minimal auxiliary assumptions to directly assess the rupture kinetics in atomic point contacts, and to compare the extracted kinetic parameters to different microscopic models.[4,5] This opens the door not only for studying bond breaking kinetics at the atomic or potentially single molecule level, but also for helping design improved nanoscale wire constructs that remain stable for extended periods of time, in turn facilitating more complex multiprobe-based break junction experiments. We show that by means of a powerful Bayesian approach we can use *all* the data and therefore the complete experimental shape of the histogram. This allows us to invert the full experimental distribution, providing robust estimates for previously inaccessible kinetic parameters even in cases where standard maximum likelihood estimation (MLE) methods fail, and leading us to a reassessment of the free energy for breaking Au-Au bonds. We begin by briefly presenting typical experimental lifetime distributions from Au break junctions and how they are influenced by externally controlled parameters. We then develop the Bayesian framework to extract kinetic parameters for bond breaking in Au junctions and validate our approach. Finally, we discuss the meaning of the thus retrieved kinetic parameters in the light of previous estimates by simpler methods.

In MCBJ experiments, an electrical bias is applied across a metallic wire. This Au wire is then stretched via a piezo motor to the point of rupture, forming a nanoscopic gap between two electrodes. The electrodes are eventually mechanically pushed back together by reversing the piezo motor extension to reform the wire, and the stretching and reforming processes are repeated. We record the conductance of these wires as a function of time during the breaking and reforming process. Fig. 1(a) shows a typical example "trace" of one stretching cycle, showing the decreasing conductance with stretching as the wire thins, followed by sudden rupture and eventually exponential decay due to tunneling through the newly formed gap. The general shape of these traces is preserved when the mechanical stretching and reforming cycle is repeated thousands of times. We visualize the aggregate data in a 2D-conductance-time histogram, shown in Fig. 1(b), clearly highlighting the characteristic behavior (decreasing conductance, then rupture followed by exponential decay) of Au junctions. Zooming in more closely into a single trace, Fig. 1(c) displays well-known quantized steps in conductance when the contact area of the wire is only a few atoms wide. Of specific interest is the amount of time that the contact conductance spends near 1 $G_0$ (the fundamental quantum of conductance), say within a window from 0.8 to 1.2 $G_0$, as shown in Fig. 1(c). This conductance window corresponds to a monovalent Au-Au contact area of an atomically thin wire just before rupture, and its lifetime describes the electromechanical properties of such an Au wire. To understand and interpret these properties, thousands of traces are acquired and their lifetimes near 1 $G_0$ are collected and binned, shown in Fig 1(d) (see SI for the effect of changing this window on the estimated parameters).

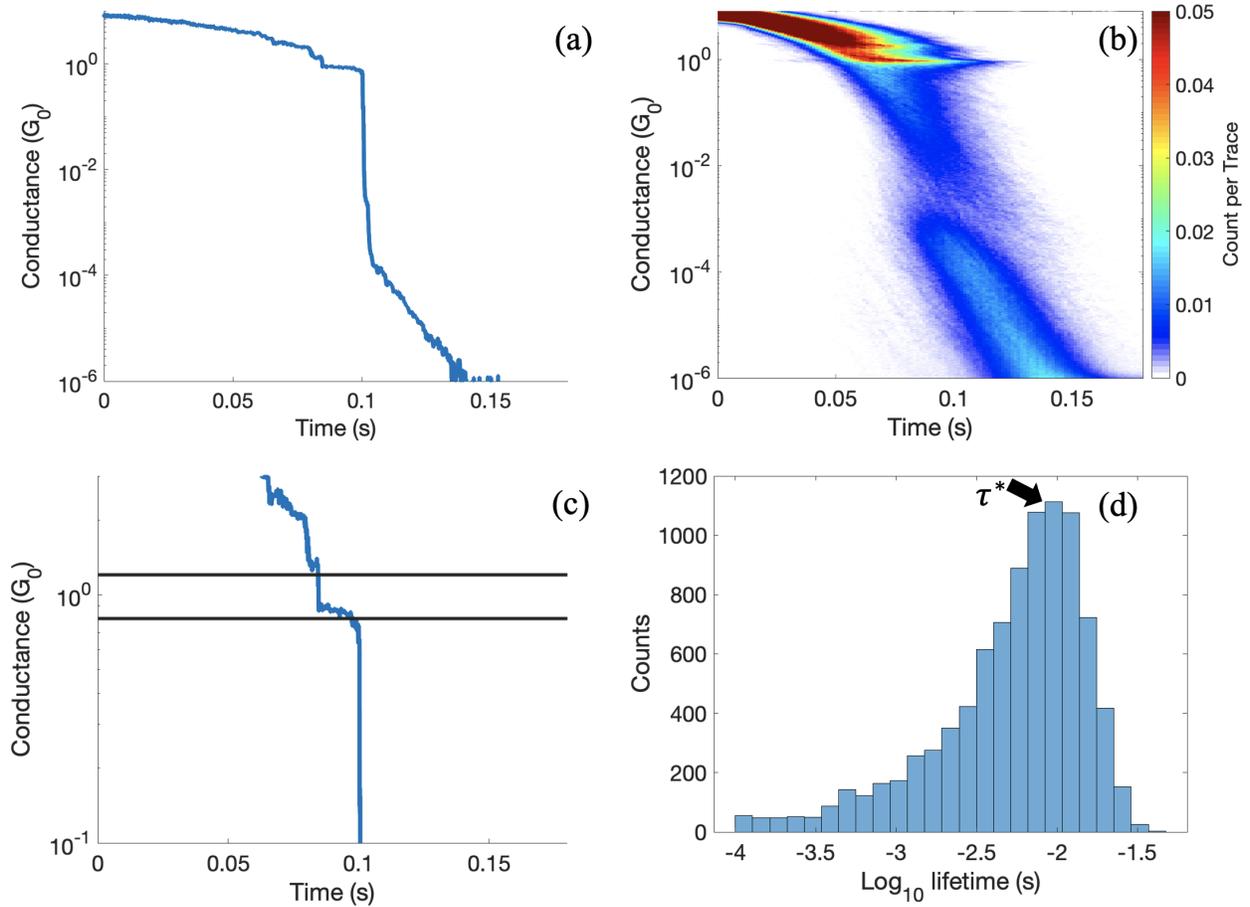

**Figure 1.** (a) Example conductance-time "trace" of one stretching cycle in our breakjunction experiment. (b) Aggregate data of thousands conductance-time "traces" in a 2D-conductance-time histogram. (c) Zoomed in conductance-time "trace" displaying the quantized steps in conductance when the contact area of the wire is only a few atoms wide, with a conductance window drawn from 0.8 to 1.2 $G_0$. (d) 1D-lifetime histogram near 1 $G_0$, with the most probable lifetime, $\tau^*$, indicated by the black arrow.

The characteristically highly asymmetric lifetime histogram may be summarized by identifying e.g. the most probable lifetime ($\tau^*$). This indicates that stable contacts of a particular configuration or at least a particular lifetime are statistically favored to form. However, for a constant stretching speed, the measured certainty in $\tau^*$ is low, apparent from observed lifetime values spanning several orders of magnitude, with a tail towards short lifetimes due to rare rupture events caused by thermal fluctuations. We attribute the shape of the lifetime distribution and the lack of certainty in $\tau^*$ to the stochastic nature of the atomic arrangement in the metallic wire during the stretching, rupture, and reforming process, and to the fact that Au atoms at room temperature are mobile.

Interestingly, the lifetime distribution shares a similar single peak structure across different stretching speeds. Despite the aforementioned shortcomings, we therefore first consider the most probable lifetime $\tau^*$ as a function of junction stretching speed (Fig. 2) as a key parameter that may encode the relevant breaking kinetics, as originally proposed by Evans and coworkers.[6,7]

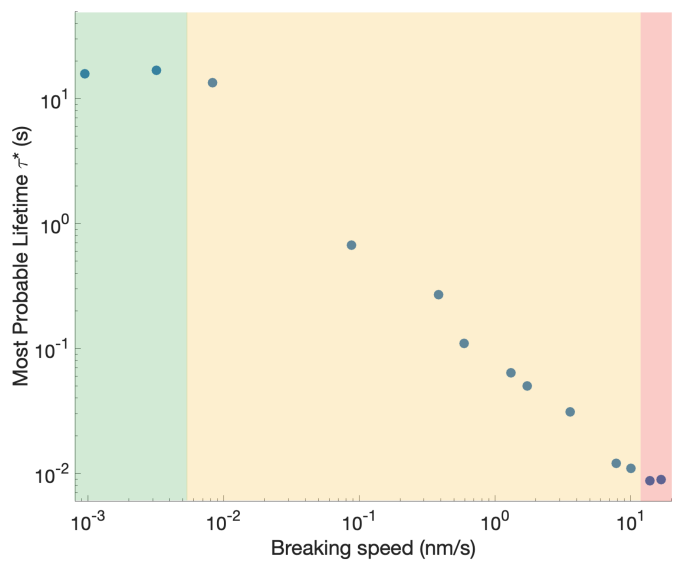

**Figure 2.** The most probable lifetime $\tau^*$ as a function of junction stretching speed. The green segment highlights spontaneous rupture due to thermal fluctuations (speed independent). The yellow segment highlights the mixed spontaneous rupture and force induced rupture (exponentially dependent on breaking speed). The red segment highlights instantaneous force induced rupture (speed independent).

At extremely slow stretching speeds, the most probable lifetime is determined by breaking events caused by spontaneous thermal fluctuations. It is independent of the stretching speed, as highlighted in the green segment of Fig. 2. At moderate stretching speeds, the most probable lifetime displays logarithmically linear behavior because of the external force increasing quickly enough to lower the free energy barrier for rupture before a spontaneous rupture can occur (yellow segment in Fig. 2). Finally, at extreme stretching speeds, bond breaking due to spontaneous thermal fluctuations is completely suppressed and rupture is instead dominated by the applied force. This force rapidly approaches the maximum tensile strength of the bond. This regime is highlighted in the red segment in Fig. 2.

While this behavior and the three different regimes is readily understood, it is difficult to obtain robust estimates of the relevant bond breaking parameters from such data. To overcome this hurdle, we take inspiration from the extensive body of work on protein bond rupture kinetics. In an effective one-dimensional microscopic model, Dudko et al. were able to capture all relevant aspects of bond breaking in a simple rate equation,[3]

$$k(F) = k_0 \left(1 - \frac{vFx^{\ddagger}}{\Delta G^{\ddagger}}\right)^{1/v-1} e^{\Delta G^{\ddagger}[1-(1-vFx^{\ddagger}/\Delta G^{\ddagger})^{1/v}]}$$

(1)

where $k(F)$ is the rate of rupture of a Au-Au bond as a function of an externally applied force $F$, $k_0$ is the intrinsic rate of a Au-Au bond breaking under zero force, $x^{\ddagger}$ is the distance from the free energy minimum to the transition state, $\Delta G^{\ddagger}$ is the height of the free energy barrier to be overcome for rupture, and $v$ is the curvature of the free energy surface. Eqn. (1) describes both constant-force and time-dependent linear force experiments, allowing both forces and lifetimes to be analyzed by the simple relation $F = KV\tau$, where $K$ is the effective spring constant and $V$ is the stretching speed. For $v = 1$ and $\Delta G^{\ddagger} \to \infty$ independent of $v$, Eqn. (1) reduces to the well-known phenomenological expression by Bell, $k(F) = k_0 e^{Fx^‡/k_b T}$, where $k_b$ is the Boltzmann constant and $T$ is temperature.[5] Bell's expression is frequently used to extract from mean rupture force measurements in optical tweezer-single molecule pulling experiments the intrinsic rate coefficient ($k_0$) and the distance along the pulling direction between the free-energy minimum and the transition state ($x^‡$).[6,7] Note that Bell's expression does not allow one to estimate the height or curvature of the free energy barrier and its surface ($\Delta G^‡, v$), which is arguably the most insightful and physically intuitive parameters in the description of the kinetics of bond breaking.

Under the assumptions of the model by Dudko et al., analytical formulae for the mean and variance of the rupture force distribution can be derived. Hummer and Szabo show that if one were to try and extract all three kinetic parameters $k_0$, $x^‡$ and $\Delta G^‡$ (when $v = \frac{1}{2}$) by fitting merely the mean forces for a range of stretching speeds from experimental data, then the $\chi^2$ surfaces are marred by highly correlated parameters with large uncertainties in the fit.[8] This situation is analogous to estimating the kinetic parameters for Au junction rupture in atomic point contact experiments. Performing a global fit of mean forces and the variances over all pulling speeds improves the situation, but extracting all three parameters, and most importantly $\Delta G^‡$, remains extremely challenging.

Instead, a statistically much more powerful and sounder approach is to use the entire rupture time *distribution*. From Eqn. (1) and considering the junction survival probability, Dudko et al obtained an analytical expression for the lifetime distribution for a constant stretching speed:[3]

$$p(\tau|V) = k(F(\tau))e^{k_0/x^‡KV} \times e^{-[k(F(\tau))/x^‡KV][1-(vKV\tau x^‡/\Delta G^‡)]^{1-1/v}}$$

(2)

Since standard MLE techniques either fail or provide poor estimates of the kinetic parameters, and since we have a readily accessible if complex parametrized stochastic model for the junction lifetimes in Eqn. (2), we show now that it is natural to adopt Bayesian estimation procedures to extract relevant electromechanical and physically meaningful junction properties. In what follows we demonstrate that such a Bayesian approach has the powerful advantage of making maximal use of *all* the measured data, i.e. unlike MLE it is not just based on most likely lifetimes or other similar single point measures. This is a central result of our work and leads to the reassessment of key kinetic parameters for bond breaking in the paradigmatic Au nanowires.

Using Eqn. (2) to describe the probability distribution of our experimentally measured lifetimes of monovalent Au-Au contacts, we invoke Bayes theorem to analytically calculate the posterior probability distribution of the most likely parameter quartet $(k_0, x^\ddagger, \Delta G^\ddagger, v)$ given our experimental data,

$$p(k_0, x^\ddagger, \Delta G^\ddagger, v \mid \tau) = \frac{p(\tau|V)p(k_0, x^\ddagger, \Delta G^\ddagger, v)}{p(\tau)}$$

(3)

where $p(\tau|V)$ is the likelihood function, i.e. the mathematical model believed to describe the data, $p(k_0, x^\ddagger, \Delta G^\ddagger, v)$ is the prior distribution which expresses our belief about the probability of the parameter quartet before any evidence is considered, and $p(\tau)$ is a normalization term. Before Bayes theorem may be used, two choices must be made. The first is that a prior distribution must be decided on. An unavoidable fact of Bayesian reasoning is that bias will *always* be introduced into the calculation of the posterior when selecting a prior distribution; we emphasize however that the explicit nature of stating the bias is one of the advantages of Bayesian statistics over frequentist analysis, where the bias is implicit and diffuse. Considering this, we quantify information of our mathematical model (Eqn. (2)) into the prior distribution by adopting the robust reference prior method developed by Berger and Bernardo (see SI for an outline of how to construct the reference prior).[9] The second choice, which is specific to our study, is how to handle $k_0$. While it would be ideal to compute the posterior for the quartet of parameters ($k_0, x^‡, \Delta G^‡, v$), it is more tractable to instead determine $k_0$ experimentally, and then estimate the triplet ($x^‡, \Delta G^‡, v$). We show in the SI that even when $k_0$ is varied by a factor of 2, it does not substantially change our estimates of the triplet.

We now use this framework to assess the electromechanical properties of a nanoscale Au wire. We have experimentally measured the inverse of the intrinsic lifetime for which an Au-Au bond will survive under zero force ($k_0$, thermally activated rupture), and set that value to be the inverse of the average of the two most probable lifetimes shown in the green region of Fig 2. We then use our Bayesian approach, including a reference prior, to estimate $x^‡, \Delta G^‡,$ and $v$ for different breaking speeds. Fig 3 shows an example of the reference prior distribution, the full posterior distribution, and the density based 95% credible interval calculated from the full posterior distribution. The credible interval expresses our uncertainty of the estimated triplet value. Note that the bounds chosen when calculating the prior distribution do not restrict the 95% credible interval, showing that our choice of bounds does not influence the estimation. This is found for all fits investigated. Table 1 summarizes the resulting 95% credible intervals calculated from the full posterior distribution for seven different breaking speeds that span approximately 2 orders of magnitude in breaking speed.

All three parameters can be reliably estimated, free of the crippling correlation effects that plague MLE approaches, and with remarkably tight credibility intervals. Both the free energy barrier and the shape of the free energy surface are largely independent of the stretching speed. This is as expected since they underpin and determine the breaking kinetics in the first place. We note that our estimates do however show a slight positive correlation of ($\Delta G^‡$) to breaking speed, varying ~ 50 meV from the center speed of 1.20 nm/s in either direction. The resulting uncertainty is however an order of magnitude lower than in previous attempts to obtain $\Delta G^\ddagger$, as discussed below. The shape of the free energy landscape ($v$) is cusp-like ($v = \frac{1}{2}$, as opposed to linear cubic, $v = \frac{2}{3}$), regardless of the breaking speed. In terms of the distance to the transition state ($x^\ddagger$), we find that it increases with breaking speed. We interpret this in the context of breakjunction experiments, where the coordination of the Au atoms sensitively determines their bond strength. The reaction coordinate is thus expected to include a sequence of complex rearrangements of the atoms in the tips to attain the most stable configuration as the wire is traveling along the lowest energy pathway, and these rearrangements are expected to depend on the applied force and thus the breaking speed (see SI for more details).

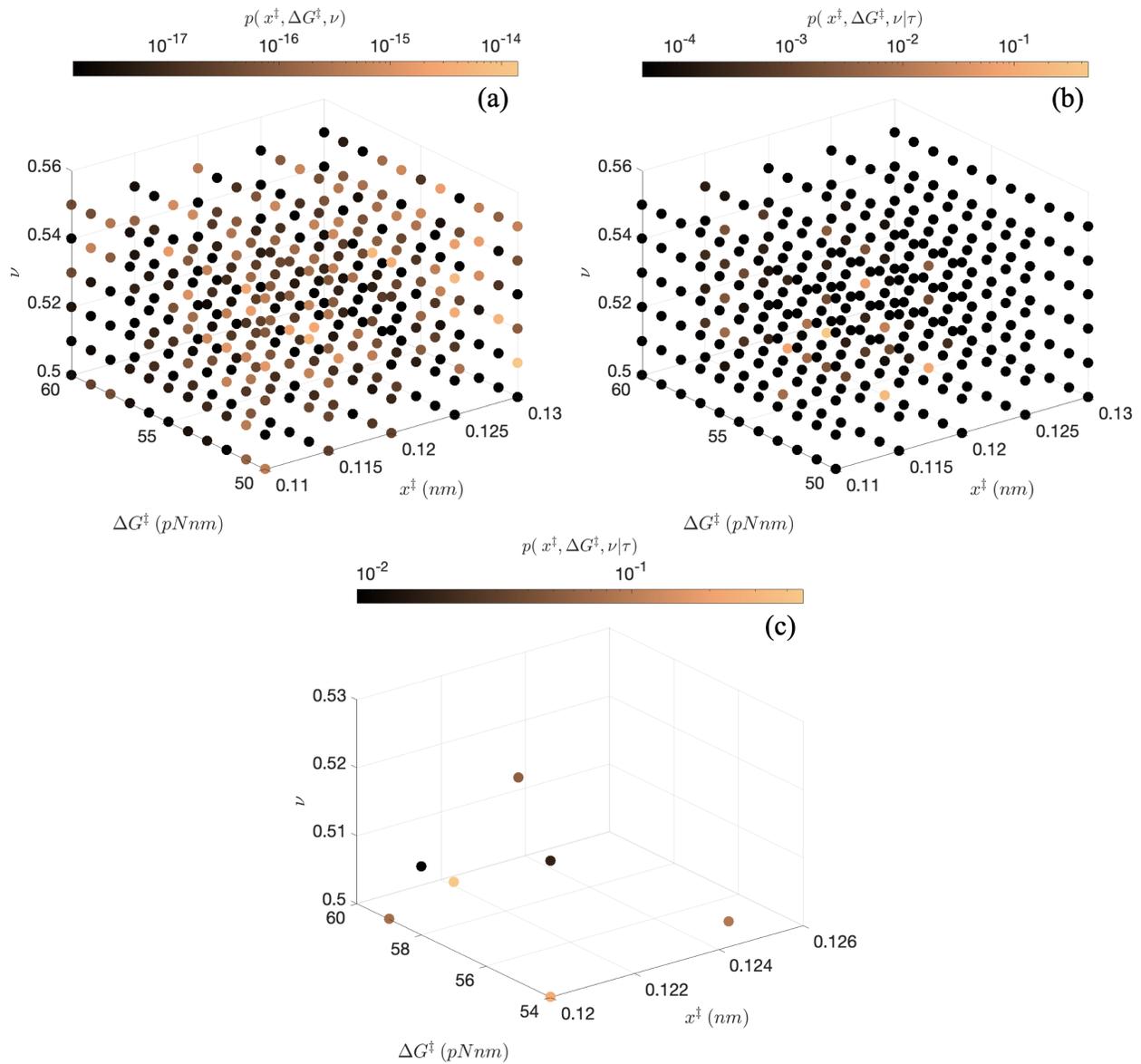

**Figure 3.** (a) The calculated reference prior distribution which quantifies the amount of expected information of Eqn. (2) into the prior distribution. (b) The full posterior distribution. (c) The calculated density-based 95% credible interval. The breaking speed is 0.65 nm/s and 160 pN nm = 1 eV.

**Table 1.** Model parameters obtained by fitting experimental lifetime data with a reference prior distribution. A single value reported rather than a range indicates that the uncertainty of the estimate in $x^{\ddagger}$ is less than 0.05 nm, and for $\Delta G^{\ddagger}$ less than 0.013 eV.

| Experimental Condition | 95% Credible Interval | | |
|---|---|---|---|
| Breaking Speed (nm/s) | $x^{\ddagger}$ (nm) | $\Delta G^{\ddagger}$ (eV) | $v$ |
| 0.09 | 0.050 | 0.294 - 0.319 | 0.52 - 0.55 |
| 0.40 | 0.075 | 0.338 | 0.50 - 0.52 |
| 0.65 | 0.120 - 0.125 | 0.337 - 0.369 | 0.50 - 0.53 |
| 1.20 | 0.120 - 0.125 | 0.350 - 0.369 | 0.50 - 0.51 |
| 3.80 | 0.120 - 0.125 | 0.381 - 0.469 | 0.50 - 0.55 |
| 11.9 | 0.145 - 0.150 | 0.375 - 0.413 | 0.50 - 0.52 |
| 15.5 | 0.135 - 0.145 | 0.406 - 0.438 | 0.50 - 0.53 |

In summary, the results of our Bayesian estimation demonstrate unambiguously that we can obtain robust estimates *of all relevant kinetic parameters* for bond rupture in atomic point contacts. Moreover, and in stark contrast to previous attempts to obtain such parameters, one need not measure many junctions with speeds distributed over many orders of magnitude, a process that takes many days for MCBJ experiments. Instead, a single speed, perhaps augmented by an independent estimate of $k_0$, suffices to retrieve the relevant kinetic parameters. This is a vast improvement on the requirements for obtaining bond rupture kinetics in single molecule transport or atomic point contact measurements, adding a new dimension to the physics that may be routinely observed in break junctions. In its essence, it is enabled by using the complete distributions rather than single point measures such as most likely values of lifetimes. In what follows, we discuss the physical meaning of the parameters retrieved and compare them to previous estimates and various theoretical models.

Despite intense interest and recent progress to go beyond average values in breakjunction experiments by understanding the shape of the complete histogram,[1,2] there remain significant challenges e.g. in extracting microscopic information from lifetime histograms. For lifetime histograms, the ms to s timescale of breakjunction measurements is significantly slower than the atomic-scale relaxation times of the junction under the stress of stretching. Because of this separation in timescales, molecular dynamics (MD) simulations cannot readily provide microscopic insight into the experimentally relevant temporal evolution of the junction during stretching, and the interpretation of the experimental lifetime distributions is thus usually limited to single point measures such as the average or most likely value. This leaves both the information content of the distribution unclear, and by necessity therefore does not reflect the full physics at play. To determine the free energy of bond breaking more accurately, we suggest that the practice of summarizing the lifetime distribution through single point measures needs to be reassessed. Instead, all the data that determines the shape of the histogram should be used to capture the physics of kinetic force-based junction rupture. It is due to this contrast that our approach can meaningfully estimate kinetic parameters from a one-dimensional model that yields the full lifetime distribution. Naturally, the underlying model assumptions require discussion. We will first carefully consider the impact of the experimental technique on the measurement. This matters, since real measurements deviate in meaningful ways from the idealized description of pulling experiments, and the consequences must be included properly. Second, we will consider alternative theoretical models that have been used in the past to understand lifetimes in breaking kinetics. Finally, we put the parameters estimated from our measurements and our approach into context with what is already known.

Arguably, the most important aspect determining lifetime is the arrangement of the atoms in and near the junction during the process of stretching the wire. It is well known from experiment that suspended Au nanowires under an external force tend to form one atom thick constrictions whose contact conductance is approximately 1 $G_0$.[10] Surprisingly, MD simulations of gold nanowires breaking under external force show the tendency to form one atom thick chains ranging from 2 - 5 atoms in length.[11,12] This peculiar behavior suggests that bonds of incompletely coordinated Au atoms are considerably stronger than fully saturated bonds in the bulk.

This is supported by previous experimental results: Rubio-Bollinger et al. performed experimental force measurements on atomic gold chains with an STM-based force sensor at 4 K, supplemented by *ab initio* calculations.[13] Remarkably, the simulations showed that the largest atomic displacements after formation of the Au chain take place in the bulk of the contact near the chain ends and not in the chain itself. Furthermore, for short chains (1-2 atoms in length), the experimental chain stiffness can vary by up to a factor of 3 due to the elasticity of the nanostructure, mainly determined by the compliance of the neighboring atoms in the bulk wire near the chain ends.

From this, we reason that there are at least two severe consequences on the measured lifetimes which will necessarily broaden their distributions. The first is the variation of Au chain length. When Au atoms near the apex are incorporated into a chain, this induces large irreversible force relaxations which must be compensated by additional stretching of the junction, as shown by Rubio-Bollinger et al. Hence, not only will the apparent breaking distance, *L*, be longer, but the time to rupture is also longer than that of the idealized junction since the latter does not need to undergo major relaxations before rupture.

The second consequence is that the stochastic geometric arrangement of the neighboring Au atoms near the nanowire ends is what determines the free energy barrier. It is clear from the exponential dependence of the breaking rate on the free energy barrier for breaking in Eqn. (1) that slight changes in this barrier height can lead to massive changes in lifetime. This reasoning is supported by calculations by Todorov et al. on gold chains, who show that a change of only 0.2 eV in the barrier will change the lifetime by 5 orders of magnitude.[14]

To properly account for these consequences of breakjunction measurements, there must be a microscopic theory of breaking kinetics that specifies the underlying free energy surface along the pulling direction to which our experimental lifetimes are sensitive. We show what kind of meaningful microscopic theory is suitable for capturing the essential aspects of breakjunction dynamics by first considering alternative theoretical models that have been used in the past to understand lifetimes in breaking kinetics and compare them to the model we suggest instead.

Kramers' treatment of chemical reactions under external bias $F$ in terms of Brownian escape from a single potential well provides a sound definition of a scalar reaction coordinate $x$ and results in a generalized Arrhenius law for the lifetime:

$$\tau = \tau_D \, e^{(\Delta G^{\ddagger} - F x^{\ddagger})/k_B T}$$

(4)

where $\tau_D$ is the diffusion relaxation time which drives thermally activated escape, $\tau$ is the measured lifetime, $\Delta G^{\ddagger}$ is the free energy barrier, and $x^{\ddagger}$ is the distance to the transition state.[15]

Kawai and coworkers showed that by measuring $\tau$ in an MCBJ at zero applied force, Eqn. (4) can be used to estimate the free energy barrier.[16] However, this result was derived with a somewhat crude approximation of the underlying free energy surface (i.e. only considering a small window to integrate near the maximum of the potential barrier, rather than integrating a full functional form). The nonlinear dependence of lifetime on free energy implies a more complex relationship between the experimentally observed parameter $\tau$ and the desired $\Delta G^{\ddagger}$. Moreover, Eqn. (4) alone cannot provide a theoretical prediction of the lifetime distribution, much less what is contributing to the shape. Though this simple model provides the basis for linking lifetime and free energy for reaction, a more comprehensive theory is needed that also captures the nature of the free energy landscape.

A more sophisticated model was proposed by Bell[5] who suggested that for the case where many bonds contribute to the barrier, the rate constant needs to be modified to include the amount of force per bond, i.e. $k_0 e^{Fx^{\ddagger}/k_B T N_b}$, where $N_b$ is the number of bonds and $k_0 = \frac{1}{\tau_D} e^{-\Delta G^{\ddagger}/k_B T}$. This provides a better reflection of junction breaking compared to that of Kramers theory, which only loosely defines the system as particle collisions. Bell provides the crucial insight that microscopic bond properties determine the macroscopic forces required to rupture.

Tao et al. took advantage of Bell's theory to estimate the natural lifetime, $\tau_{off} = \frac{1}{k_0}$ in an STM-BJ experiment on Au and molecular junctions.[17,18] However, while the power of Bell's equation is its generality, the expression is only valid for diffusive barrier crossing in the limit of small forces, restricting it to a limited range of breaking scenarios. Furthermore, Bell's expression does not consider the influence of a well-defined free energy surface that is coupled to the harmonic pulling potential, focusing instead only on $x^{\ddagger}$. Consequently, the underlying free energy surface is insufficiently characterized by only a single parameter, $x^{\ddagger}$, leaving the height of the free energy barrier, $\Delta G^{\ddagger}$, impossible to estimate without reverting to simplistic models or including ad-hoc assumptions.

In comparison to the theory developed by Dudko et al. which *does* specify a potential energy surface, and which underpins our Bayesian estimation of breaking kinetics, it will become clear just how important this specification is when the parameters estimated from our measurements and our approach are put into context with previous estimates. Following Kawai et al. and using Eqn. (4) to estimate the free energy barrier when $F = 0$, $\tau_D = 3.5 \times 10^{12}$ s, and transforming the measured Au lifetime distribution into a free energy barrier distribution yields a distribution centered at ~ 0.80 eV with a half width of ~ 0.1 eV.[16] Alternatively, one may use $E_b = 1/2 F_b L$, where $E_b$ is the energy barrier, $F_b$ is the maximum amount of force required to break a Au-Au bond, and $L$ is the breaking distance over which the bond can be stretched from the equilibrium bond length before breakdown. $F_b$, measured by an Atomic Force Microscope (AFM) on gold chains, is $1.5 \pm 0.3$ nN.[13] $L$, which is analogous to the distance to the transition state, $x^\ddagger$, may be determined by measuring the breaking length at extreme stretching speeds in either an MCBJ or STM-BJ setup. It is broadly distributed with a most probable value of ~ 0.17 nm. Note that the Au-Au breaking distance, $L$, also defined as the distance between the equilibrium bond length and the stretched bond length just before rupture, has been reported in Transmission Electron Microscope (TEM)[19-21], MCBJ[16,22,23], STM-BJ[17] experiments, MD[12], *ab initio* molecular dynamics (AIMD),[24] and tight binding molecular dynamics simulations[25] of Au chains breaking to have a range of 0.02 - 0.30 nm. Using the entire breaking distance distribution reported by Tao et al yields $E_b = 0.82 \pm 0.45$ eV.[17]

The uncertainty in the above estimates of the free energy barrier is rather large and is due to several factors: First, the lifetime and breaking length distributions are broadened by both gold chaining and the stochastic geometric arrangement of the Au atoms in the nanowire, which directly governs the free energy barrier. Second, the statistical analysis of the broadened lifetime and breaking length distributions, based on a theoretical treatment which does not account for the specific nature of the free energy surface, compounds the uncertainty in the estimates. As a result, the free energy barrier is systematically overestimated and burdened with large uncertainties.

In contrast, our model, though one-dimensional, requires fewer assumptions and is based on a treatment that specifies the underlying free energy surface. This is justified since experimental lifetimes are quite sensitive to the fact that the free energy surface needs to be specified by more than just $x^‡$. Most importantly, and as a major innovation of this work, we combine this with a powerful Bayesian approach to make maximal use of *all* the measured data to provide robust estimates where standard MLE techniques fail. This allows us to reliably invert experimental data and retrieve previously inaccessible or widely uncertain parameters for the kinetics of bond rupture.

In conclusion we have shown that to extract relevant electromechanical and physically meaningful junction properties from lifetime histograms, summarizing the lifetime distribution through single point measures needs to be reassessed. Instead, all the data and thus the full shape of the lifetime histogram should be used to capture the physics of junction breaking in atomic point contact measurements. When considering the impact of the MCBJ experimental technique on the measurement of Au-Au lifetimes carefully, two new necessary innovations are called for when analyzing the lifetime data: First, a theoretical treatment that specifies the underlying free energy surface by more than just $x^‡$. This is needed because experimental break junction data are sensitive to the stochastic geometric arrangement of the neighboring Au atoms near the nanowire ends, which is what determines the free energy barrier. Second, the adoption of Bayesian fitting procedures which makes maximal use of *all* the measured data. This is crucial since standard MLE techniques that rely on single point measures and/or binning the data can lead to overestimation with large uncertainties or may even completely fail, as in the case of fitting the experimental lifetime distribution to Eqn. (2).

These lessons apply more broadly to many other experimental approaches that rely on interpretation of broadly distributed stochastic data. Clearly, our approach can be used to assess bond rupture dynamics in single molecule transport measurements. Applications to force measurements in biophysics also benefit from the proposed framework. Finally, we expect that entirely different physical scenarios with inherently stochastic data that yield broad, asymmetric distributions and for which analytical models of the distribution may be developed, will be amenable to the Bayesian approach in our work.

**Supporting Information**.

Additional complications and consequences of the breakjunction architecture, testing the Bayesian methodology, experimental methods.

ATHOUR INFORMATION

*Dylan Dyer* - Department of Chemistry and Biochemistry, The University of Arizona, Tucson, Arizona 85721, United States.

*Oliver L.A. Monti* - Department of Chemistry and Biochemistry, The University of Arizona, Tucson, Arizona 85721, United States; Department of Physics, The University of Arizona, Tucson, Arizona 85721, United States. Phone: +520 626 1177, Email: monti@arizona.edu

**Notes**

The authors declare no competing financial interests.

ACKNOWLEDGMENTS

The authors would like to acknowledge support from the National Science Foundation award no. DMR-2225369, as well as the Alfred P. Sloan Foundation award no. G-2020-12684. Plasma etching was performed in part using a Plasmatherm reactive ion etcher acquired through an NSF MRI grant, award no. ECCS-1725571, as well as an AGS reactive ion etcher located in the Micro/Nano Fabrication Center at the University of Arizona. Bayesian statistical calculations were performed using High-Performance Computing (HPC) resources supported by the University of Arizona TRIF, UITS, and RDI and maintained by the UA Research Technologies department. Quality control was performed using a scanning electron microscope in the W. M. Keck Center for Nano-Scale Imaging in the Department of Chemistry and Biochemistry at the University of Arizona with funding from the W. M. Keck Foundation Grant. A special thanks goes to Koen Fischer and Chris Jarzynski for their wonderful collaborative spirit and invaluable guidance on theoretical background in force spectroscopy. We also thank Manny Smeu and Hashan Peiris for insightful discussions.

# Supporting Information

# Bond Breaking Kinetics in Mechanically Controlled Break Junction Experiments: A Bayesian Approach

*Dylan Dyer[1] and Oliver L.A. Monti[1,2]\**

[1] Department of Chemistry and Biochemistry, The University of Arizona, Tucson, Arizona 85721, United States

[2] Department of Physics, The University of Arizona, Tucson, Arizona 85721, United States

\* Phone: +520 626 1177, Email: monti@arizona.edu

**Contents**

# S1. Additional complications and consequences of the breakjunction architecture

## *S1.1 Effect of the breaking speed distribution*

Before actual experimental data can be analyzed, an additional complication, a consequence of the architecture of any break junction experiment, must be accounted for. Interpreting conductance-distance data from a Mechanically Controlled Break Junction (MCBJ) requires the conversion of change in piezo motor distance ($\Delta z$) to change in inter-electrode distance ($\Delta d$) through the attenuation ratio $r = \frac{\Delta d}{\Delta z}$. The stretching speed for each cycle of Au-wire rupture is then calculated by $V = \frac{\Delta d}{\Delta t}$. Here, $\Delta t$ is the amount of time it takes the piezo motor to travel $\Delta z$ in the upward direction, which is attenuated to $\Delta d$ in the three-point bending beam geometry of MCBJs.[1] In our experimental setup, we determine this attenuation ratio by fitting the slope of the tunneling current in piezo motor distance, as is standard in the field of breakjunction physics.[2] When the attenuation ratio is calculated in this way, there will *always* be a distribution of attenuation ratios for any given sample, because the tunneling current depends sensitively on the stochastic atomic configuration of the metallic wire during the stretching, rupture, and reforming process. Consequently, the observed lifetime distributions are broadened by the variation of inter-electrode stretching distance for a given piezo displacement and breaking speed during each stretching cycle, and this effect must be included to model the observed lifetime distributions. Similar considerations also apply to Scanning Tunneling Microscope-Break Junctions (STM-BJ). Hence, to capture an experimental lifetime histogram such as the one shown in Fig. 1(d) of the main manuscript, we must modify Eqn. (2) in the main manuscript to be a weighted sum over individual instances of stretching speed in each respective stretching cycle.

$$p(\tau) = \sum_{i=1}^{N} p(\tau|V_i)p(V_i)$$

(SI.1)

where *N* is the total number of different stretching speeds for the given dataset.

We first show an example of a breaking speed distribution calculated as described above. Each tunneling trace in the 2D-histogram shown in Fig SI.1. (a) is individually fit to determine the attenuation for the given trace. We bin the thousands of attenuation ratios and display them in a 1D-histogram as shown in Fig SI.1. (b). These attenuation ratios are then converted into inter-electrode breaking speeds (Fig SI.1. (c)). The plots are shown for data collected at 60 $\mu$m/s.

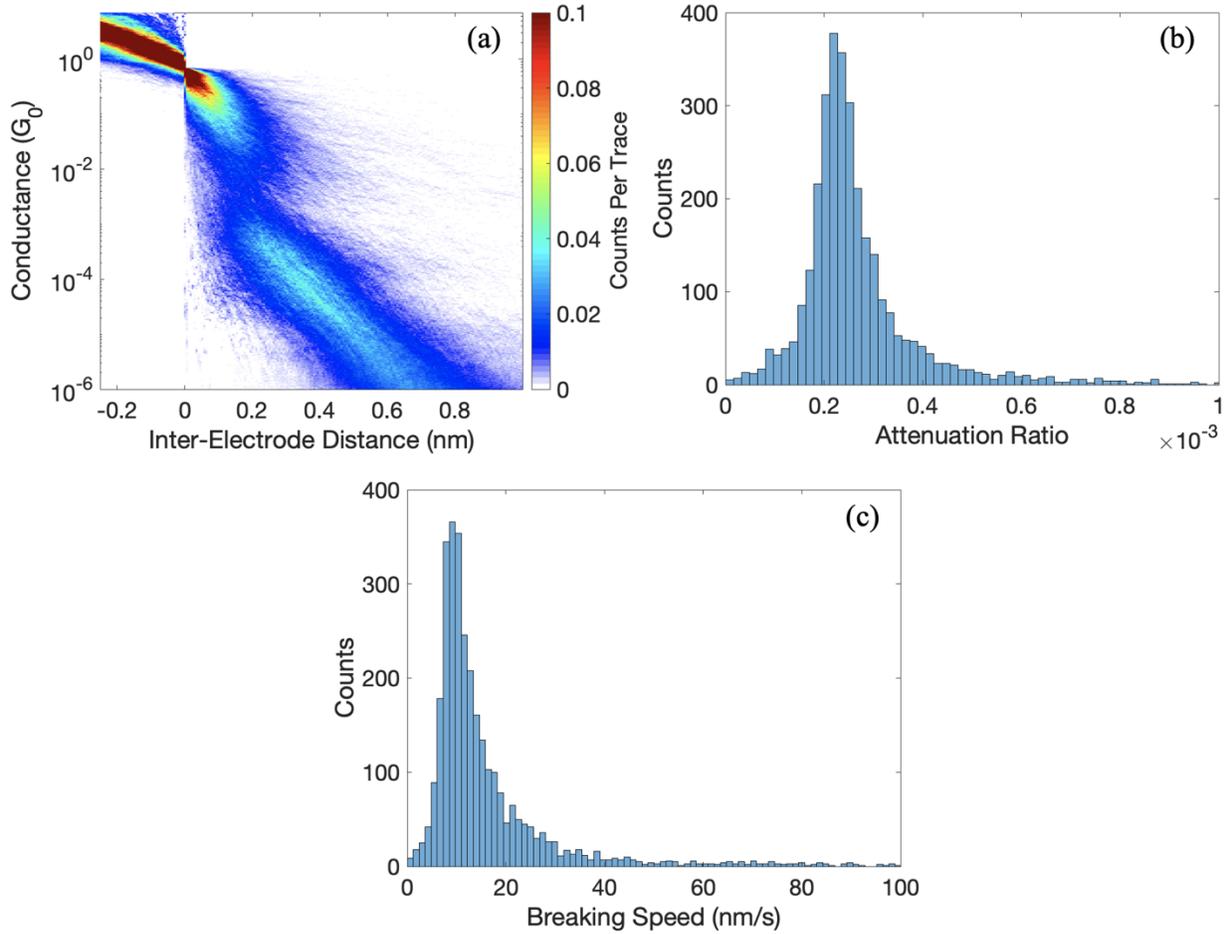

**Fig. SI.1.** (a) Conductance-distance tunneling traces binned into a 2D-histogram. (b) 1D-histogram of attenuation ratios, each from an individual breaking trace. (c) 1D-histogram of breaking speeds calculated from the attenuation distribution.

We now show in Fig. SI.2. what exactly this modification does by comparing histograms created from data sampled from the modified expression (SI.1) and unmodified expression in the main manuscript (Eqn. 2). First, we expect that the breaking speed distribution will not change the single point measures of the unmodified distribution such as the mean or most probable values. This is true because the shape of the breaking speed distribution shows a clear peak structure at a preferred speed, which is the speed that will be used in the unmodified analytical expression.

Second, we expect that since we are experimentally sampling from an ensemble of breaking speed scenarios which closely resemble the average or most probable speed, our lifetime distributions will be broadened.

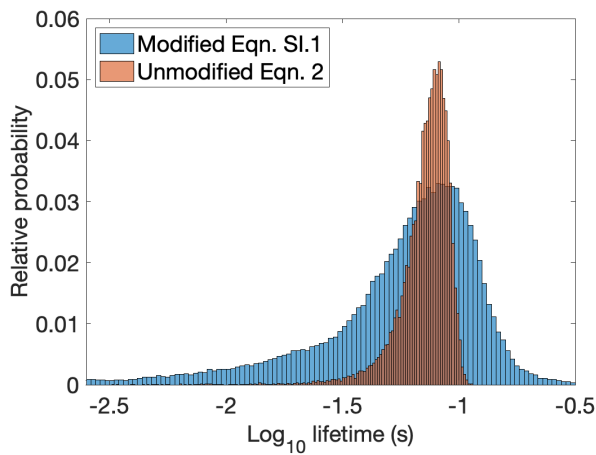

**Fig. SI.2.** Overlaid 1D-histograms of data sampled from the modified equation and the unmodified equation. As expected, the mean values remain very close at 0.069 s and 0.073 s for the modified and unmodified data that was sampled. The modified equation produces more widely dispersed lifetimes due to sampling from many different speeds at their relative occurrence.

*S1.2 Dependence of $x^{\ddagger}$ on breaking speed*

To obtain appropriate statistics in the context of the model proposed by Dudko et al., hundreds of copies of putatively identical molecular structures must be analyzed. In biological systems where the types of bonds, loosely defined as the number of participating bonds, their strength, and their coordination to each other may for the most part be always the same, the distance to the transition state is not expected to depend strongly on the breaking speed ($V$). However, in an MCBJ experiment the coordination of the Au atoms is stochastic. This sensitively determines the individual bond strength between each pair of atoms. The statistics we sample are from a collection of different Au junction orientations rather than a single mostly unchanging bonding arrangement as the model assumes. Therefore, the reaction coordinate is expected to include a sequence of complex rearrangements of the atoms in the tips to attain the most stable configuration as it is traveling along the lowest energy pathway. We hypothesize that these rearrangements are expected to depend on the applied force and thus the breaking speed such that, $x^{\ddagger} \sim F \sim f(V)$.

From Dudko et al. it is known that the qualitative behavior of $F$ is predicted to be $\langle F \rangle \sim \ln(V)^{\nu}$, and hence $x^{\ddagger} \sim \langle F \rangle \sim \ln(V)^{\nu}$. Since the breaking distance, $L$, is linearly related to $F$ by $F = kL$, where $k$ is the spring constant, and because $L$ is an observable from our experiment, we substitute $L$ for $F$ such that $x^{\ddagger} \sim \langle L \rangle \sim \ln(V)^{\nu}$ and compare their qualitative behavior. As shown in Fig. SI.3., the qualitative behavior of $x^{\ddagger}$ and $L$ is very similar and follows the predicted behavior. This explains the observed speed dependence of the extracted value of $x^{\ddagger}$ discussed in the main manuscript and in the following sections. Note that we choose the median of $L$, rather than the average because it is a better measure of centrality in our case.

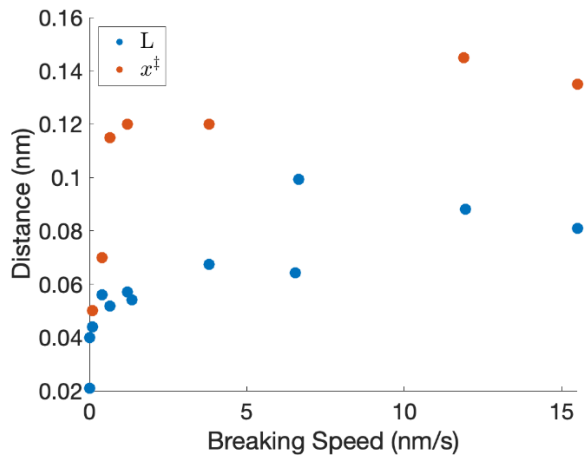

**Fig. SI.3.** Plot showing predicted behavior of the median of measured breaking lengths, *L*, and the estimated distance to the transition state ($x^\ddagger$) from the fits in the main manuscript.

## S3. Testing the Bayesian methodology

### S3.1 Reference Prior Method

For each parameter of interest in a model ($\theta_1, \theta_2, \theta_3 \ldots$), choose the range over which they span. Choose the first numerical value for each parameter of interest. Then compute the Fisher Information with respect to the last parameter of interest,

$$I(\theta_3) = \int_{\mathbb{R}} \left( \frac{\partial}{\partial \theta_3} \log f(x|\theta_1, \theta_2, \theta_3) \right)^2 f(x|\theta_1, \theta_2, \theta_3) \, dx$$

(SI.2)

then compute the conditional prior point $\pi(\theta_3|\theta_2)$,

$$\pi(\theta_3|\theta_2) = \frac{\exp\left[\frac{1}{2} \log I(\theta_3)\right]}{\int \exp\left[\frac{1}{2} \log I(\theta_3)\right] d\theta_3}$$

(SI.3)

Now compute the Fisher information for the second parameter of interest,

$$I(\theta_2) = \int_{\mathbb{R}} \left( \frac{\partial}{\partial \theta_2} \log f(x|\theta_1, \theta_2, \theta_3) \right)^2 f(x|\theta_1, \theta_2, \theta_3) \, dx$$

then compute the conditional prior point $\pi(\theta_2|\theta_1)$,

$$\pi(\theta_2|\theta_1) = \frac{\exp\left[\frac{1}{2}\log I(\theta_2)\right]}{\int \exp\left[\frac{1}{2}\log I(\theta_2)\right] d\theta_2}$$

(SI.5)

Now compute the Fisher information for the final parameter of interest

$$I(\theta_1) = \int_{\mathbb{R}} \left(\frac{\partial}{\partial \theta_1} \log f(x|\theta_1, \theta_2, \theta_3)\right)^2 f(x|\theta_1, \theta_2, \theta_3) \, dx$$

(SI.6)

finally, compute the first conditional prior point $\pi(\theta_1)$

$$\pi(\theta_1) = \exp\left\{-\frac{1}{2}\iint \pi(\theta_3|\theta_2)\pi(\theta_2|\theta_1) \log I(\theta_1) \, d\theta_2 d\theta_3\right\}$$

(SI.7)

The final reference prior point is then defined as,

$$\pi(\theta_3, \theta_2, \theta_1) = \pi(\theta_3|\theta_2)\pi(\theta_2|\theta_1)\pi(\theta_1)$$

(SI.8)

## S3.2 Variation of conductance window

Our standard choice of conductance window is from 0.8 to 1.2 $G_0$. This agrees with other studies of the properties of monovalent Au chains.[3-8] The lifetime distribution with the standard conductance window, as shown in Fig. SI.4. (a), shows a clear peak structure with a tail towards short lifetimes due to the stochastic nature of the atomic arrangement in the metallic wire during the stretching, rupture, and reforming process, and because Au atoms at room temperature are mobile. When the conductance window is closed to a value of 0.9 to 1.1 $G_0$, there is significant loss of shape in the distribution, particularly at short lifetimes, as shown in Fig. SI.4. (b). Upon opening the conductance window to a value of value of 0.7 to 1.3 $G_0$, the shape of the distribution is preserved, with perhaps a slight shift in the peak towards higher lifetime values (Fig SI.4. (c)).

Table SI.1. summarizes the extracted relevant kinetic parameters from each distribution, withholding the distribution created from a conductance window of 0.9 to 1.1 $G_0$ since it could not be fit, likely due to incomplete capture of the full histogram. The differences in the estimated parameters from the two distributions that could be fit are small in the case of ($x^‡$), while ($\Delta G^‡$) and ($\nu$) is nearly identical.

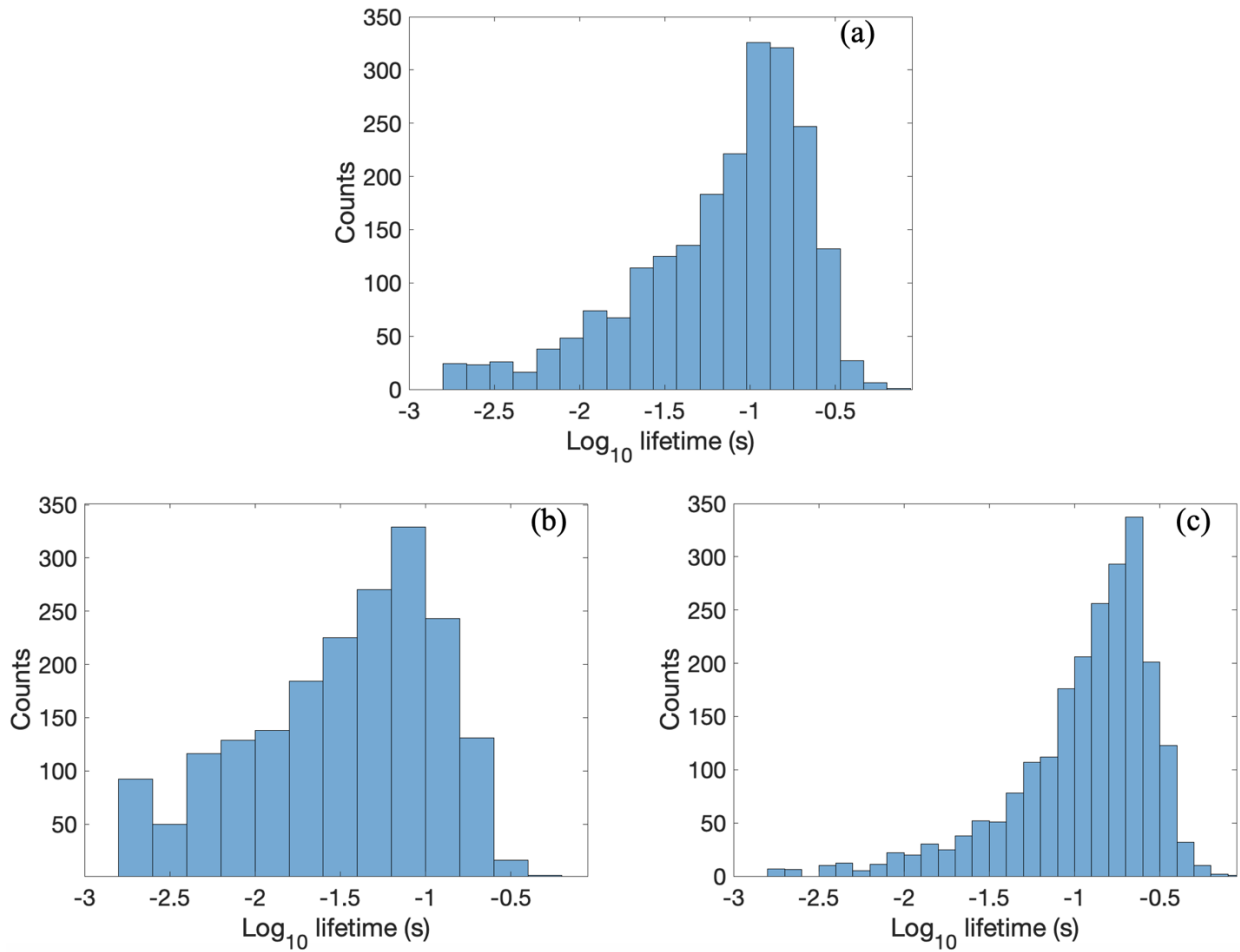

**Fig. SI.4.** (a) 1D-histogram showing the lifetime distribution with a conductance window of 0.8 to 1.2 $G_0$. (b) 1D-histogram showing the lifetime distribution with a conductance window of 0.9 to 1.1 $G_0$. (c) 1D-histogram showing the lifetime distribution with a conductance window of 0.7 to 1.3 $G_0$.

**Table SI.1.** Model parameters obtained by fitting experimental lifetime data for $v$ as a fitting parameter with a standard prior, and varying the conductance window. A single value reported rather than a range indicates that the uncertainty of the estimate in $x^{\ddagger}$ less than 0.05 nm.

|  | 95% Credible Interval | 95% Credible Interval |
| --- | --- | --- |

| Experiment condition | $G_{window}$ = 0.80 to 1.20 | | | $G_{window}$ = 0.70 to 1.30 | | |
|---|---|---|---|---|---|---|
| Breaking Speed (nm/s) | $x^‡$ (nm) | $\Delta G^‡$ (eV) | $v$ | $x^‡$ (nm) | $\Delta G^‡$ (eV) | $v$ |
| 0.65 | 0.115 - 0.125 | 0.331 - 0.369 | 0.50 - 0.55 | 0.090 | 0.344 - 0.375 | 0.50 - 0.56 |

## S3.3 Variation of $k_0$

In the main manuscript before we analyze our experimental data, we discuss the choice of how to handle $k_0$. The intrinsic rate ($k_0$) may be measured at extremely slow stretching speeds which are roughly less than 0.008 nm/s. To account for variations in where the inverse of the most probable lifetime $1/\tau^* = k_0$ might be measured, we vary the intrinsic rate by a factor 2 such that the measured $\tau^*$ may be anywhere between 8 and 33 seconds at speeds slower than 0.008 nm/s. This range of $\tau^*$ is in accordance with previous measurements performed by Kawai et al.[3] Table SI.2. shows that this choice of $k_0$ does not significantly change our estimated parameters.

**Table SI.2.** Model parameters obtained by fitting experimental lifetime data for $v$ as a fitting parameter with a standard prior, and varying $k_0$ by a factor 2. A single value reported rather than a range indicates that the uncertainty of the estimate in $\Delta G^‡$ less than 0.031 eV.

| Experiment condition | 95% Credible Interval | 95% Credible Interval | 95% Credible Interval |
|---|---|---|---|
| | $k_0$ = 0.0305 | $k_0$ = 0.061 | $k_0$ = 0.122 |

| Breaking Speed (nm/s) | $x^{\ddagger}$ (nm) | $\Delta G^{\ddagger}$ (eV) | $\nu$ | $x^{\ddagger}$ (nm) | $\Delta G^{\ddagger}$ (eV) | $\nu$ | $x^{\ddagger}$ (nm) | $\Delta G^{\ddagger}$ (eV) | $\nu$ |
|---|---|---|---|---|---|---|---|---|---|
| 0.65 | 0.135 - 0.140 | 0.375 | 0.50 - 0.54 | 0.115 - 0.125 | 0.331 - 0.369 | 0.50 - 0.55 | 0.100 - 0.105 | 0.313 - 0.344 | 0.50 - 0.54 |

## S3.4 Effect of setting $\nu$ vs. fitting $\nu$ on experimental data

We explored two ways to determine the curvature of the free energy surface. First, we computed the posterior distribution and its respective 95% density based credible interval for $\nu = \frac{1}{2}$ and for $\nu = \frac{2}{3}$ respectively and compared the two ("fitting method 1"). "Fitting method 2" treats the parameter $\nu$ as a random variable while fitting. Table SI.3. shows a comparison of the two methods.

For method 1, we see the largest discrepancies between the distance to the transition state ($x^{\ddagger}$) at breaking speeds of 0.65 nm/s and 1.20 nm/s, while the differences in the values of ($x^{\ddagger}$) at speeds of 0.09, 0.40, 3.80, 11.9, and 15.5 nm/s are reasonably small. The largest discrepancies for the free energy barrier ($\Delta G^{\ddagger}$) are at breaking speeds of 1.20 nm/s and 11.9 nm/s, with all other speeds resulting in a barrier that is essentially the same. The somewhat arbitrary fluctuations between the two set values of $\nu$ can make this method potentially unreliable.

For method 2, the distance to the transition state increases with breaking speed, as discussed in the main manuscript. The free energy barrier is largely independent of the breaking speed with a slight positive correlation, and the curvature resembles a cusp-like shape ($v = \frac{1}{2}$) that is independent of the breaking speed.

When comparing the two methods, we thus consider only ($v = \frac{1}{2}$) for method 1 since it more closely resembles the freely fit values of $v$. For the distance to the transition state ($x^{\ddagger}$), there is a clear positive correlation with breaking speed in fitting method 2 (freely fit $v$). While the positive correlation in fitting method 1 is also observed, it is noisier. In contrast, for the free energy barrier ($\Delta G^{\ddagger}$), there is little difference between the estimates. However, most importantly, considering the overall trends of both methods when $v$ is cusp-like, ($x^{\ddagger}$) is positively correlated with speed and ($\Delta G^{\ddagger}$) is largely speed-independent. Given that this is true for both methods, and given that the values of $x^{\ddagger}$ vary more systematically in method 2, we consider method 2 preferable.

**Table SI.3.** Model parameters obtained by fitting experimental lifetime data for $v = \frac{1}{2}, v = \frac{2}{3}$, and $v$ as a fitting parameter. Fits were performed with a reference prior distribution. A single value reported rather than a range indicates that the uncertainty of the estimate in $x^{\ddagger}$ is less than 0.05 nm, and in $\Delta G^{\ddagger}$ less than 0.013 eV.

| | Fitting Method 1 | | | | Fitting Method 2 | | |
|---|---|---|---|---|---|---|---|
| Experimental condition | 95% credible interval | | 95% credible interval | | 95% credible interval | | |
| | $v = 1/2$ | | $v = 2/3$ | | | | |
| Breaking Speed (nm/s) | $x^{\ddagger}$ (nm) | $\Delta G^{\ddagger}$ (eV) | $x^{\ddagger}$ (nm) | $\Delta G^{\ddagger}$ (eV) | $x^{\ddagger}$ (nm) | $\Delta G^{\ddagger}$ (eV) | $v$ |
| 0.09 | - | - | 0.07 | 0.375 | 0.05 | 0.294 - 0.319 | 0.52 - 0.55 |
| 0.40 | 0.137 - 0.148 | 0.300 - 0.331 | 0.129 - 0.138 | 0.300 - 0.331 | 0.075 | 0.338 | 0.50 - 0.52 |
| 0.65 | 0.116 - 0.124 | 0.331 - 0.363 | 0.174 - 0.186 | 0.338 - 0.363 | 0.120 - 0.125 | 0.337 - 0.369 | 0.50 - 0.53 |
| 1.20 | 0.122 - 0.126 | 0.344 - 0.356 | 0.165 - 0.170 | 0.238 - 0.243 | 0.120 - 0.125 | 0.350 - 0.369 | 0.50 - 0.51 |
| 3.80 | 0.171 - 0.179 | 0.350 - 0.356 | 0.158 - 0.164 | 0.350 - 0.363 | 0.120 - 0.125 | 0.381 - 0.469 | 0.50 - 0.55 |
| 11.9 | 0.218 - 0.224 | 0.300 - 0.306 | 0.172 - 0.179 | 0.344 - 0.350 | 0.145 - 0.150 | 0.375 - 0.413 | 0.50 - 0.52 |
| 15.5 | 0.190 - 0.196 | 0.350 - 0.356 | 0.172 - 0.176 | 0.356 - 0.363 | 0.135 - 0.145 | 0.406 - 0.438 | 0.50 - 0.53 |

## S3.5 Effect of standard prior vs reference prior on experimental fits

Here we provide a comparison of the estimated parameters when a default prior is used vs. when a reference prior is used. A default prior is one that assumes all outcomes are equally likely (a uniform distribution). As shown in Table SI.4., the general case is that the reference prior improves the density based 95% credible interval.

**Table SI.4.** Model parameters obtained by fitting experimental lifetime data for $v$ as a fitting parameter with both a standard and reference prior. A single value reported rather than a range indicates that the uncertainty of the estimate in $x^{\ddagger}$ is less than 0.05 nm, and in $\Delta G^{\ddagger}$ less than 0.013 eV.

| Experimental condition | 95% credible interval | | | 95% credible interval | | |
|---|---|---|---|---|---|---|
| | Standard Prior | | | Reference Prior | | |
| Breaking Speed (nm/s) | $x^{\ddagger}$ (nm) | $\Delta G^{\ddagger}$ (eV) | $\nu$ | $x^{\ddagger}$ (nm) | $\Delta G^{\ddagger}$ (eV) | $\nu$ |
| 0.09 | 0.05 | 0.294 - 0.319 | 0.50 - 0.55 | 0.05 | 0.294 - 0.319 | 0.52 - 0.55 |
| 0.40 | 0.070 - 0.075 | 0.325 - 0.350 | 0.50 - 0.55 | 0.075 | 0.338 | 0.50 - 0.52 |
| 0.65 | 0.115 - 0.125 | 0.331 - 0.369 | 0.50 - 0.55 | 0.120 - 0.125 | 0.337 - 0.369 | 0.50 - 0.53 |
| 1.20 | 0.125 | 0.350 | 0.50 | 0.120 - 0.125 | 0.350 - 0.369 | 0.50 - 0.51 |
| 3.80 | 0.120 - 0.125 | 0.375 - 0.481 | 0.50 - 0.55 | 0.120 - 0.125 | 0.381 - 0.469 | 0.50 - 0.55 |
| 11.9 | 0.145 - 0.150 | 0.375 - 0.413 | 0.50 - 0.52 | 0.145 - 0.150 | 0.375 - 0.413 | 0.50 - 0.52 |
| 15.5 | 0.135 - 0.145 | 0.406 - 0.438 | 0.50 - 0.53 | 0.135 - 0.145 | 0.406 - 0.438 | 0.50 - 0.53 |

*S3.6 Effect of standard prior vs reference prior on synthetics fits*

We test our Bayesian methodology by sampling synthetic data for a known parameter triplet $(x^{\ddagger}, \Delta G^{\ddagger}, \nu)$ from Eqn. (2) in the main manuscript via a Metropolis-Hastings algorithm and attempting to retrieve the simulated values. To test the importance of bias, we use two different methods to construct the prior distribution: Table SI.5. shows a comparison of a default prior distribution where all outcomes are equally likely, and a reference prior distribution, along with their calculated respective density-based 95% credible intervals. The credible interval expresses our uncertainty of the estimated triplet value. In both cases, the parameter triplet ($x^\ddagger = 0.40$ nm, $\Delta G^\ddagger = 0.688$ eV, $v = 0.55$) which was used to generate the synthetic data is retrieved inside of the 95% credible interval, clearly validating our Bayesian approach. We emphasize that standard maximum likelihood estimation (MLE) methods based on most likely lifetimes as used previously struggle with retrieving such triplets and are plagued by considerable parameter correlation and ensuing large uncertainties. In contrast, the Bayesian approach presented here is quite robust to the choice of prior and retrieves the parameter triplet with good fidelity, regardless of the simulation parameters. The advantage of using a reference prior is that the amount of expected information (i.e. what the parameter triplet ($x^\ddagger, \Delta G^\ddagger, v$) tells about the lifetime, $\tau$, and its certainty, $p(\tau|V)$, *before* any experimental observation) is quantitatively introduced in the prior distribution.

**Table SI.5.** Model parameters for lifetime distributions obtained by fitting simulated lifetime data with different prior distribution choices. A single value reported rather than a range indicates that the uncertainty in the estimate in $x^\ddagger$ is less than 0.1 nm.

| Prior Choice | 95% Credible Interval | | |
|---|---|---|---|
| | $x^\ddagger$ (nm) | $\Delta G^\ddagger$ (eV) | $v$ |
| Uniform Prior | 0.40 | 0.625 - 0.731 | 0.51 - 0.61 |
| Reference Prior | 0.40 | 0.625 - 0.743 | 0.50 - 0.61 |

## S3. Experimental methods

MCBJ samples were fabricated on a substrate of 0.5 mm thick phosphor bronze coated with a few micron thick insulating layer of polyimide. The pattern of a thin metal wire with an ~

100 nm constriction in the center was defined using electron beam lithography. The wire itself was created by thermally evaporating a 4 nm titanium adhesion layer followed by 80 nm of gold. Finally, reactive ion etching with an $O_2/CHF_3$ plasma was used to turn the central constriction into an ~ 1 μm free-standing gold bridge by under etching the polyimide. MCBJ experiments were performed in air at room temperature. Each MCBJ sample was cleaned with $O_3$/UV and rinsed with ethanol shortly before use. For each sample, we initially deposited ~ 10 μL of pure dichloromethane, using a clean glass syringe, on the center of the junction, and confined over the bridge with the aid of a Kalrez gasket. Each sample was clamped into a custom-built three-point bending apparatus in which a push rod was used to bend the sample and thereby thin and break the gold bridge. To collect each breaking trace, a stepper motor was first used to adjust the push rod until the conductance through the gold bridge was between 5 and 7 $G_0$ (where $G_0$ is the quantum of conductance, equal to 77.48 μS). When this set point was reached, the collection of a single breaking trace was triggered by raising the push rod 40 μm at a given piezo speed using a linear piezo actuator while simultaneously recording the conductance through the bridge at 20 kHz using a custom high-bandwidth Wheatstone bridge amplifier.[9] The piezo was then retracted, after which the process was repeated to collect additional breaking traces. Custom LabVIEW software was used to automatically collect thousands of consecutive breaking traces. During trace collection, the bending apparatus was placed on a vibrationally isolated table to reduce mechanical noise and inside a copper Faraday cage to reduce electrical and acoustic noise. We collect a tunneling data set of a few thousand breaking traces and calculate the attenuation ratio of each respective trace from the MCBJ sample.